# Acceleration-guided Acoustic Signal Denoising Framework Based on Learnable Wavelet Transform Applied to Slab Track Condition Monitoring

Baorui Dai, Gaëtan Frusque, Qi Li, and Olga Fink

**Abstract**—Acoustic monitoring has recently shown great potential in the diagnosis of infrastructure conditions. However, due to the severe noise interference in acoustic signals, meaningful features tend to be difficult to infer. It creates a considerable obstacle to an extensive application of acoustic monitoring. To tackle this problem, we propose an acceleration-guided acoustic signal denoising framework (AG-ASDF) based on learnable wavelet transform to automatically denoise the acoustic signal and extract the relevant features based on the acceleration signal. This denoising framework requires the acceleration signal only for the training stage. Therefore, only acoustic sensors (nonintrusive) need to be installed during the application phase, which is convenient and crucial for the condition monitoring of safety-critical infrastructure. A comparative study is conducted among the proposed AG-ASDF and other feature learning/extraction methods, by using a multiclass support vector machine (SVM) to evaluate the detection effectiveness of slab track conditions based on acoustic signals. Different healthy and unhealthy states of slab tracks are imitated with three types of slab track supporting conditions in a railway test line. The classification based on the proposed AG-ASDF features outperforms other feature extraction and learning methods with a significant accuracy improvement.

**Index Terms**— Acceleration signal, acoustic signal, condition monitoring, deep learning, denoising, slab track.

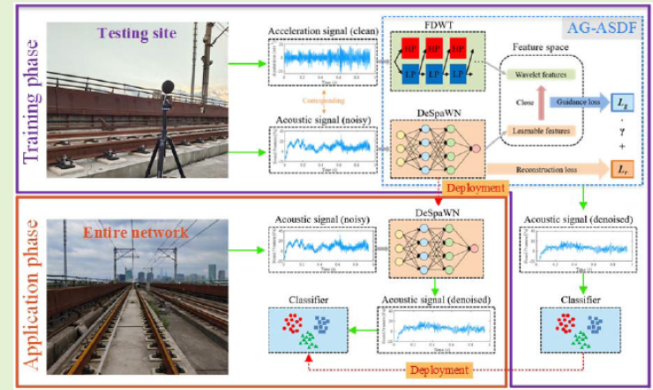

## I. Introduction

THE implementation of concrete slab track solutions has been recently increasing particularly for high-speed railway lines because of their advantages in operation and maintenance [1]. Influenced by the environmental and operating conditions such as dynamic train load and temperature fluctuations, the mortar layer, which is serving as connection between slab track and foundation, is prone to degradation. The degradation of mortar layer will weaken the supporting condition of slab track, thus, affecting the comfort of passengers and even the safety of trains [2, 3]. Different degrees of mortar layer degradation will lead to different maintenance actions. Particularly detecting the very early degradation conditions may help to prolong the useful lifetime of the mortar layer. Thus, the classification of different slab track states is a very important step in the maintenance decision-making process.

Acceleration sensors have been one of the preferred choices in structural health monitoring systems to monitor the condition of railway infrastructure [4, 5]. The technology has the advantage of directly reflecting the vibration characteristics of the measured objects. However recently, some concerns have been emerging about the application of accelerometers due to the potential risks to the safety of railway operations stemming from the close proximity to railway clearance placement of sensors [6].

Recently, monitoring the condition of industrial and infrastructure assets with acoustic signals has been gaining importance since acoustic sensors are non-intrusive and easy to install or retrofit [7-9], which is crucial for safety-critical infrastructure like railways. Although promising results have been obtained on the general industrial assets monitoring [10, 11], there have been few studies addressing the application of acoustic monitoring in railways. For example, Pieringer et al. [12] adopted a noise measurement car to collect wheel-rail noise and trained a logistic regression classifier to identify squats in the German railway network. In another research study, Wang et al. [13] used a smartphone to collect

This study was supported by the National Natural Science Foundation of China (grant numbers 52178432 and 51878501) and China Scholarship Council (grant number 202106260178). (Corresponding author: Qi Li.)

Baorui Dai is with the Department of Bridge Engineering, Tongji University, Shanghai 200092, China, and also with the Laboratory of Intelligent Maintenance and Operations Systems, EPFL, Lausanne 1015, Switzerland (e-mail: dbr@tongji.edu.cn).

Gaëtan Frusque and Olga Fink are with the Laboratory of Intelligent Maintenance and Operations Systems, EPFL, Lausanne 1015, Switzerland (e-mail: gaetan.frusque@epfl.ch; olga.fink@epfl.ch).

Qi Li is with the Department of Bridge Engineering, Tongji University, Shanghai 200092, China, and also with the Tibet Agriculture & Animal Husbandry University, Nyingchi 860000, Tibet, China (e-mail: liqi_bridge@tongji.edu.cn).

acceleration and sound signals inside subway trains, and applied the extreme gradient boosting algorithm to identify squeal and rumble as a reflection of rail defection. Meng et al. [14] used an optical fiber sensing system to pick up the acoustic signals along the fiber line and achieved a high-precision detection of railway perimeter intrusions. In general, research and application of acoustic condition monitoring for railways has been recently evolving. However, the condition monitoring of slab tracks based on acoustic signals has remained unaddressed.

In this research, we aim to perform a proof of concept and to evaluate the detectability and distinguishability of the different potential degradation states of the mortar supporting layer based on acoustic signals. Due to the random dynamic excitation and environmental factors, the acoustic signals measured from the field monitoring are non-stationary [7]. Time-frequency spectrograms and wavelet coefficients are considered to be effective in capturing meaningful features of non-stationary signals [15, 16]. However, careful feature extraction with the right hyperparameter selection is required to extract a compact representation of the raw high-frequency signals. Moreover, the effectiveness of the extracted features for condition identification might be affected by unexpected noise or changing operating conditions. To address these limitations, a deep learning framework named Denoising Sparse Wavelet Network (DeSpaWN) was recently proposed, which can automatically learn meaningful and sparse representations of raw high-frequency signals [8]. The DeSpaWN achieves very good results compared to other state-of-the-art approaches on acoustic anomaly detection and classification tasks.

Although having the advantage of non-intrusiveness, acoustic monitoring signals are often impacted by severe noise which creates a considerable obstacle for an extensive application of acoustic monitoring. Compared to acoustic signals, acceleration signals are much less impacted by noise because accelerometers are in contact with the surface of the measured objects so that the target vibration can be captured directly. Acoustic signals and acceleration signals are both the products of structural vibration. Hence, there is a potential relationship between the two kinds of signals [15]. Motivated by this fact and the promising results of DeSpaWN in previous research studies, we develop an acoustic denoising framework that can learn a strong denoising transform under the guidance of the corresponding acceleration signals. Although the acoustic denoising framework requires simultaneously acceleration and acoustic signals to train a deep learning model, only the acoustic signals are later used in the application phase. In a practical setup this would correspond to the case that both signals are collected on a testing site and later in the rollout to the entire network, only acoustic monitoring devices would be installed.

To evaluate the performance of the proposed acoustic denoising framework, a railway field monitoring experiment is conducted in this study. One of the typical deterioration states of slab tracks is caused by degraded mortar layer, which changes the supporting condition of the slab track. In other words, the connection between track slab and foundation becomes weaker. Since abnormal states of slab tracks in railway lines are rare in real applications, we aim to collect data from a railway test line by imitating the degraded condition of the mortar layer by using different types of support layers. More concretely, this study uses three types of slab track with different supporting conditions: mortar, rubber and spring support in a railway test line. These three conditions correspond to the healthy state, intermediate degradation, and severe degradation condition. Acceleration sensors (contact) and acoustic sensors (contactless), are installed next to the three types of slab track to collect the acceleration and acoustic signals as a train passes by with different speeds. Based on the acoustic signals (acceleration signals are only required in the training phase of AG-ASDF), a comparative study on the detection effectiveness of slab track condition is conducted among different feature learning / extraction methods, by applying a multi-class support vector machine (SVM) on the extracted / learning features.

The remainder of this paper is arranged as follows: In Section II, the architecture of the AG-ASDF, the field experiment setup, and the classification method are introduced. In Section III, classification results are obtained and discussed to conduct the comparative study. In Section IV, conclusion and future works are presented.

## II. DATA AND METHODS

A flow chart of the content of this study is illustrated in Fig. 1. In the following subsections, the data collection scheme is first introduced. Then we present the specific theories of DeSpaWN and AG-ASDF in sequence. To conduct a comparative study on the effectiveness of the time-frequency representation, we adopt other advanced feature learning / extraction methods to obtain the comparable features. Finally, a classification method is introduced to validate the feature learning performance of the AG-ASDF compared to other methods.

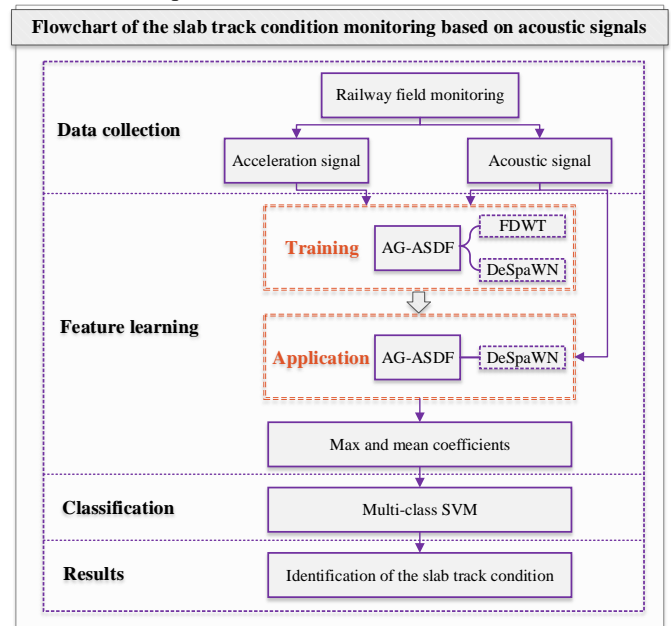

Fig. 1. Flowchart of slab track condition monitoring.

### A. Data collection

We use three types of slab track with different supporting conditions in a railway test line to imitate different healthy and degraded states of slab tracks with different levels of degradation, as shown in Fig. 2. The connections between the three track slabs and foundations are mortar (concrete), rubber, and discrete spring support, respectively. We consider mortar layer as the healthy condition of the support layer, rubber layer as intermediate degradation level of the support layer and the spring support as degraded support layer. This results, therefore, in three classes of health conditions: one healthy class and two classes of degraded conditions with two different severity levels.

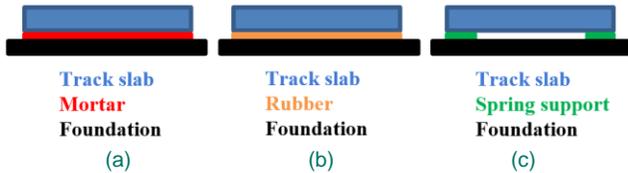

Fig. 2.  Supporting conditions of three types of slab track: (a) No degradation; (b) Intermediate degradation; (c) Severe degradation.

The train operated on the railway test line is a metro train, with six vehicles and a total length of 140 m. Acceleration sensors and acoustic sensors are installed next to the three types of slab track to collect the acceleration and acoustic signals as the train passes with different speeds (20, 40, 60, and 80 km/h), as shown in Fig. 3. The relative positions of acceleration sensors, acoustic sensors and the test slab tracks are the same for the three types of slab track. The sampling frequencies of the acceleration sensors and acoustic sensors are 20 kHz. The collected dataset of each track type consists of 24 samples (the train passes six times under each speed) for both acceleration and acoustic signals.

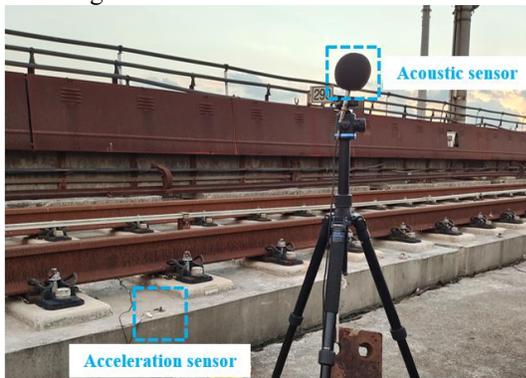

Fig. 3.  Placement of acoustic and acceleration sensors.

To facilitate the subsequent feature extraction, we limit the effective duration of each signal between the first and the last wheel sets of the moving train. The average effective durations under different train speeds are presented in Table I. Slab Track 1, 2, and 3 represent the healthy condition, intermediate degradation level, and severe degradation level of support layers, respectively.

TABLE I
AVERAGE EFFECTIVE DURATIONS OF COLLECTED SIGNALS

| Speed (km/h) | Number of train passes | Average duration of intercepted signals (s) | | |
|---|---|---|---|---|
| | | Slab Track 1 | Slab Track 2 | Slab Track 3 |
| 20 | 6 | 21.84 | 22.34 | 22.04 |
| 40 | 6 | 11.41 | 11.47 | 11.43 |
| 60 | 6 | 7.94 | 7.85 | 7.90 |
| 80 | 6 | 6.11 | 6.12 | 6.20 |

## B.  Denoising Sparse Wavelet Network

Recently, several architectures that combine the interpretability advantages of signal processing and the learning capabilities of neural networks have shown promising results [17-20]. The DeSpaWN [8], a recently proposed deep learning framework inspired by fast discrete wavelet transform (FDWT), appears to be particularly adapted for our task as it decomposes the input signal in different and adapted time-frequency resolutions.

Fig. 4 illustrates the cascade algorithm related to the FDWT, which forms the basic architecture of DeSpaWN. A low-pass and a high-pass filter which are both followed by a sub-sampling step decompose an input signal into detail and approximation coefficients. Recurrently, the approximation coefficients of the previous layer are decomposed in a similar way. The detail coefficients have accurate time-frequency resolution varying according to the layer. By including the approximation coefficients of the last layer, they form the time-frequency representation of the input signal. From the obtained representation, it is possible to perfectly reconstruct the input signal via the inverse FDWT.

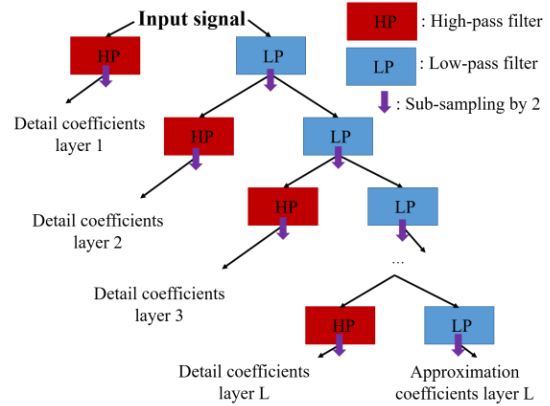

Fig. 4.  Architecture of the cascade algorithm related to the FDWT. L corresponds to the selected number of layers.

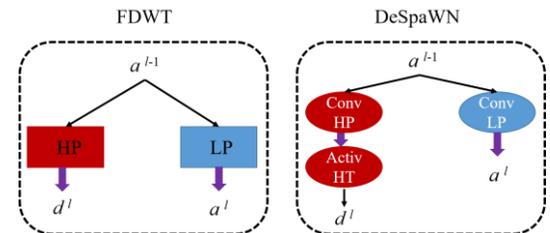

Fig. 5.  Encoding blocks of the FDWT and the DeSpaWN. $a^{l-1}$, $a^l$ and $d^l$ are the approximation coefficients of the ($l$-1) layer, approximation coefficients of the $l$ layer, and the detail coefficients of the $l$ layer, respectively. Oval elements with white texts in the Figure are learnable.

DeSpaWN has an encoder-decoder architecture based on the successive use of the FDWT and the inverse FDWT. It utilizes a fully learnable variation of the cascade algorithm (Fig. 4) by allowing learning the kernel common to both filters at each layer. Moreover, the resulting detail coefficients (plus the approximation coefficients of the last layer) are fed to a specifically designed learnable hard thresholding (HT) activation functions. The encoding blocks of the FDWT and the DeSpaWN are shown in Fig. 5. The HT activation function is expressed as a combination of two opposite sigmoid functions [8]:

$$HT(x) = x\left[\frac{1}{1+\exp(\alpha(x+b_-))} + \frac{1}{1+\exp(-\alpha(x-b_+))}\right] \quad (1)$$

where $\alpha$ is a "sharpness" factor and is set to 10 in this paper to simulate a "sharp" threshold, $b_+$ and $b_-$ are the positive and negative learnable bias acting as the thresholds on both sides of the origin.

Similar to the FDWT, the decoding blocks of DeSpaWN have a reverse architecture compared to the encoding blocks but without the HT activation functions. Independently for each layer, the learnable HT functions operate as an automatic wavelet denoising operation in the encoding blocks inducing sparsity in the final time-frequency representation. However, due to the HT activation function, it is not possible anymore to perfectly recover the input signal with the decoding part of DeSpaWN.

To achieve firstly a good signal reconstruction and secondly a sparse decomposition, DeSpaWN uses a loss function to minimize the reconstruction error plus a sparse regularization term applied to the time-frequency representation [8], which is expressed as:

$$L = |s - \hat{s}|_1 + \gamma\left(\sum_{l \in [1,L]}|d^l|_1 + |a^L|_1\right) \quad (2)$$

where the $s$ and $\hat{s}$ are the original signal and the reconstructed signal respectively, $\gamma$ is the weight of the sparse regularization term, $l$ and $L$ represent the decomposition layer number and the maximum decomposition layer number respectively, $d$ and $a$ are the detail coefficients and the approximation coefficients respectively.

For DeSpaWN, we initialize the learnable filters with Daubechies-4 wavelets in this study, and the learnable bias of the HT activation functions with 0.5. The regularization parameter is set to 1. The DeSpaWN is trained using the Adam optimiser with a learning rate of 0.0001. Meaningful features are extracted from the learned wavelet coefficients and then used as input to the classification algorithm, which will be introduced in Sections II-D and II-E.

### C. Acceleration-guided Acoustic Signal Denoising Framework

In real applications, the acoustic signal radiated by the target object is often impacted not only by the environmental noise but also by other sounds radiated by irrelevant structures or mechanisms. As a consequence, the effective acoustic signal may be heavily impacted by non-informative noise and artifacts.

The denoising performance of unsupervised machine learning methods like DeSpaWN tends to be limited in this case.

Supervised deep denoising models [21], using a training set composed of pure signals and their noisy counterparts, have recently shown remarkable performances. In [22], an extension of DeSpaWN to the wavelet packet transform was successfully applied for background noise removal. However, in our application, it is not possible to record the pure acoustic signals of the slab track. Instead, we propose to consider the measured acceleration signals, as their signal to noise ratio (SNR) is much higher than that of the measured acoustic signals.

We then consider that the features provided by the FDWT of the acceleration signal can be seen as the target of our learnable transform applied to the acoustic signals. To enforce similarity between the acoustic and acceleration signal representations in the wavelet feature space, we developed a new framework called AG-ASDF. The architecture of AG-ASDF is shown in Fig. 6.

Since acceleration signals are more intrusive, their application may not be scalable to the entire railway network. We, therefore, assume that both signals are collected on the testing site. Later, during the application phase (and the rollout on the entire railway network), only the non-intrusive acoustic monitoring signals are available. Therefore, the input signals to AG-ASDF are different for the training and application phases. For the training stage, the simultaneously measured acceleration and acoustic signals are used as input to the AG-ASDF. The acceleration signal is passed through FDWT to be decomposed into wavelet coefficients, while the acoustic signal is fed into DeSpaWN to produce learnable features. The same number of decomposition layers is applied for both the FDWT and the DeSpaWN. In the application phase, only the measured acoustic signals are used as input to the trained AG-ASDF to obtain meaningful features. It is worth noting that the relative positions of acoustic sensors and test objects should be consistent during the training and application phases to achieve optimal denoising performance.

The proposed loss is a trade-off between the reconstruction of the acoustic signal with the DeSpaWN, and the similarity between the wavelet features and the learnable features. In a nutshell, we replace the unsupervised sparse regularization of the original DeSpaWN loss function [8] with a supervised regularization to enforce enhanced features guided by the acceleration signal. The loss function of the AG-ASDF is proposed as follows:

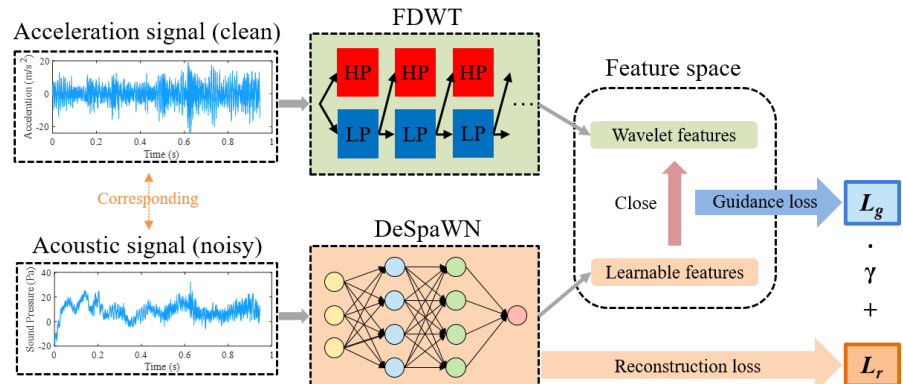

Fig. 6. Architecture of AG-ASDF. Both acoustic signals and corresponding acceleration signals are required in training stage. Only acoustic signals are required in application phase.

$$L = |s - \hat{s}|_1 + \gamma \left( \sum_{l \in [1,L]} \sum_{m \in [1,M]} \left| \text{fea}^m \left( d_{acou}^{\ l} \right) - \text{fea}^m \left( d_{acce}^{\ l} \right) \right|_1 \right. \\ \left. + \sum_{m \in [1,M]} \left| \text{fea}^m \left( a_{acou}^{\ L} \right) - \text{fea}^m \left( a_{acce}^{\ L} \right) \right|_1 \right) \quad (3)$$

where $m$ and $M$ are the feature type number and the maximum feature type number, $d_{acou}$ and $d_{acce}$ are the detail coefficients of acoustic and acceleration signals, $a_{acou}$ and $a_{acce}$ represent the approximation coefficients of acoustic and acceleration signals, the fea( ) represents the feature extraction of the wavelet coefficients.

In this work, we choose maximum and mean values of each decomposition layer of absolute wavelet coefficients as the features described in Eq. (3) to enforce the similarity between acoustic and acceleration signals from the perspective of local and global characteristics. Therefore, the maximum feature type number $M$ in Eq. (3) is set to 2 and fea( ) can be expressed as:

$$\text{fea}^1 (\ ) = \max | \ | \quad (4)$$
$$\text{fea}^2 (\ ) = \text{mean} | \ | \quad (5)$$

### D. Feature learning and extraction

In the time dimension, either an acceleration signal or an acoustic signal can be characterised by six periodical waveforms due to the successive excitation of six vehicles of the train. To enrich the dataset, each signal is divided into six time bands with equal width, and the split signals are regarded as independent sample units.

The dataset of acoustic signals is divided into training and test data with a 3:1 ratio. The six time series of the same train ride are either all in the training dataset or all in the test dataset. AG-ASDF, DeSpaWN, FDWT with 'db4' basis, Wavelet Packet Transform (WPT) with 'db4' basis, Short-time Fourier transform (STFT), Mel spectrogram, synchrosqueezing transform based on STFT (SST_STFT) [23], synchrosqueezing transform based on continuous wavelet transform (SST_CWT) [24], variational mode decomposition (VMD) [25], and scattering transform (ST) [26] are used in this study to extract features. Each data sample is padded with zeros to form an identical length of 85,404 points, which equals the maximum number of points of the split signals. Therefore, the number of layers that are decomposed by AG-ASDF, DeSpaWN and FDWT is set to $\log_2 85,404 \approx 16$. To maintain consistency between the different methods, the number of filter bands of the Mel spectrogram and the number of decomposition layers of the WPT is also set to 16. The spectrograms of STFT, SST_STFT, and SST_CWT are manually divided into 16 frequency bands with equal width. The window sizes (with overlap ratio of 1/2) for Mel spectrogram STFT, and SST_STFT are determined as 1024 to get a frequency resolution of around 20 Hz. Once the transformation is completed, the zeros at the end of each decomposition layer of signals are removed. Since it is less effective in decomposition to generate intrinsic mode functions (IMFs) of VMD as many as 16, we set the number of IMFs as a moderate value 6. The maximum scale of the low-pass filters and the number of wavelets per octave of ST are empirically set as $2^8$ and 4 according to the length and oscillation property of the measured signals. Then the scattering coefficient matrix are also manually divided into 16 bands with equal width in frequency dimension.

Generally, the maximum and mean values of the time-frequency or wavelet coefficients (converted to decibels) from each filtered frequency band of split signals are taken as the features. For DeSpaWN and AG-ASDF, there are two additional features, namely the maximum and average values of the residuals between the reconstructed and the original signal. For VMD, the features are the maximum and mean absolute values of the IMFs in time domain. For ST, the maximum and mean values of the scattering coefficients (converted to decibels) from each divided band of matrix are obtained as the features.

### E. Classification

The training and test data for classification are consistent with the training and test data for DeSpaWN as well as for AG-ASDF. A multi-class SVM with Radial Basis Function kernel is used in this study to classify the slab track conditions. In the training stage, the 5-fold cross validation technique is utilized to obtain the optimal hyperparameters of the SVM. In the application phase, the classification results are obtained by applying the SVM with optimal hyperparameters to the test data.

## III. RESULTS AND DISCUSSION

### A. Considered classification tasks

Two classification tasks are implemented to evaluate the effectiveness of slab track condition monitoring based on the acoustic signals as well as the performance of the proposed framework. For the first task, training dataset and test dataset contain all train speeds. For the second classification task, we evaluate the generalization ability and study the slab track condition when the train speed in the test dataset is different from those observed in the training dataset. To do this, the collected signals with three different train speeds are assigned to the training dataset, and the signals with the remaining speed form the test dataset.

If not stated otherwise, the results obtained with either AG-ASDF or DeSpaWN are calculated under the condition that the two parts of loss function are weighted equally. For feature learning methods, all experiments are run five times and the mean and standard deviation values of classification accuracy are reported. For feature extraction methods, since the whole process is deterministic, we run all experiments only one time and report the classification accuracy.

As a preliminary study, the classification performance using acceleration signals is evaluated. The FDWT is used to extract the features. The results show that the classification accuracy based on the acceleration signals is 100% in all classification tasks. Since the acceleration sensors are in direct contact with slab tracks, they can measure the vibration of the track structure without being impacted by the environmental interference. The classification using acoustic signals always performs worse compared to the acceleration signals. The results will be presented in detail in the following sections. It should be noted that this preliminary study motivated the development of the proposed AG-ASDF framework in which the acceleration signals are used to guide the denoising of the acoustic signal.

TABLE II
COMPARATIVE RESULTS BASED ON ACOUSTIC SIGNALS IN CLASSIFICATION TASK 1

| Slab track condition | Classification accuracy (%) | | | | | | | | | |
|---|---|---|---|---|---|---|---|---|---|---|
| | AG-ASDF | DeSpaWN | FDWT | WPT | STFT | Mel | SST_STFT | SST_CWT | VMD | ST |
| No degradation (Class 1) | **97.2**±0.4 | 91.7±1.6 | 83.3 | 88.9 | 88.9 | 77.8 | 94.4 | 88.9 | 80.6 | 88.9 |
| Intermediate degradation (Class 2) | 93.1±1.7 | 89.8±2.3 | 83.3 | 41.7 | 83.3 | 91.7 | 91.7 | **94.4** | 58.3 | 91.7 |
| Severe degradation (Class 3) | 95.8±1.3 | 98.1±0.6 | 97.2 | 66.7 | 75 | 94.4 | 80.6 | 97.2 | 44.4 | **100.0** |
| Average | **95.4** | 93.2 | 87.9 | 65.8 | 82.4 | 88.0 | 88.9 | 93.5 | 61.1 | 93.5 |

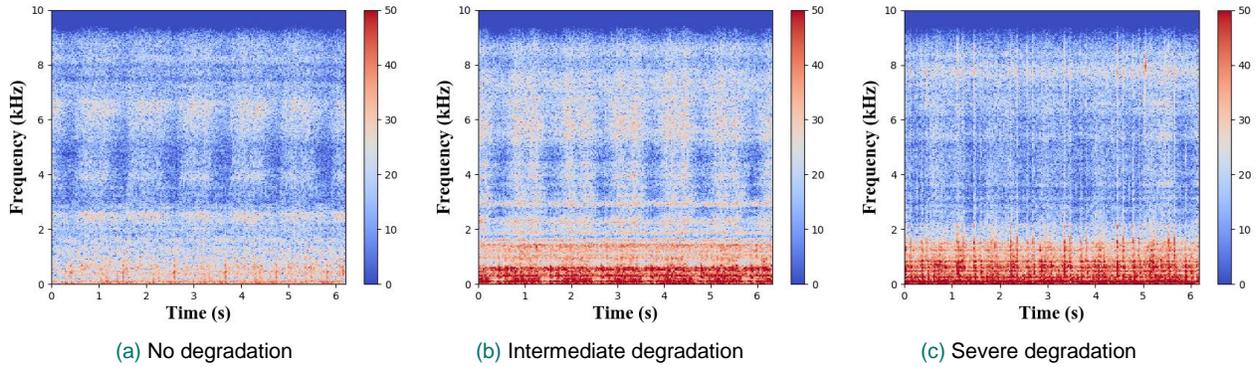

(a) No degradation  (b) Intermediate degradation  (c) Severe degradation
Fig.7. STFT spectrograms of acceleration signals under the train speed of 80 km/h.

### B. Same train speeds in training and test dataset (Task 1)

The results of the first task evaluating the classification accuracy are presented in Table II. The best performance of identifying the slab track condition based on acoustic signals does not quite reach the classification accuracy based on acceleration (95.4% on average compared to the 100% based on acceleration signals). This difference in performance can be attributed to the higher noise level of acoustic signals compared to the acceleration signals.

Table II also shows that the AG-ASDF reaches a performance improvement between 1.9% and 34.3% compared to other feature learning / extraction methods on average. This demonstrates the strong and directed acoustic signal denoising capacity of the proposed AG-ASDF facilitated by the DeSpaWN learning capability and the supervision by the acceleration signals with lower noise levels. The original DeSpaWN, SST_CWT, and ST methods almost uniformly result in the second best performance. This confirms the observations in a previous publication that DeSpaWN has the capability of learning a noise-independent representation of signals. Moreover, it shows that the ability of meaningful feature extraction possessed by SST_CWT and ST methods is excellent. WPT and STFT are the two methods that perform the worst in this condition monitoring task. This can be explained by the fact that the dominant frequency range of the slab track vibration is often below 2 kHz (Fig. 7), while the two methods give equal weights to low and high frequency features within the range of 0-10,000 Hz. On the contrary, the FDWT and the Mel spectrogram, which focus more on the characterization of the low-frequency-part of the signals, achieve a better performance compared to WPT and STFT in this case.

### C. Generalization ability: different train speeds in training and test dataset (Task 2)

In real applications of the track state monitoring task, the collected dataset for training may not contain all the train speed conditions. This puts forward the requirement to predict the slab track state under the condition of a train speed that may not have been observed during the data collection period and may, therefore, not be part of the training dataset. To imitate such a task, signals with three different train speeds are used to train the SVM classifier and the signals with the remaining speed are used to test it. In the following, the different resulting setups are introduced and obtained results are discussed.

*1) Training dataset: 20, 40, 60 km/h; Test dataset: 80 km/h*

This setup results in fact in an extrapolation task, requiring to extrapolate the learned features to higher train speeds where the noise impact is quite different compared to the lower train speeds. Table III displays the classification results based on acoustic signals with different feature leaning / extraction methods. The discrepancy between the data in the training test dataset leads to a significant decrease in the classification accuracy compared to task 1 where samples from all considered speeds were contained in the training and test datasets.

Table III also shows that the wavelet-based methods (DeSpaWN, FDWT, WPT, SST_CWT, and ST) are generally superior to the other methods (STFT, Mel spectrogram, SST_STFT, and VMD) in this task leading to more robust results. A possible explanation is that the wavelet-based methods have more flexible time and frequency resolutions at different decomposition layers compared to the other two methods, while the speed may influence the activation time of the characteristic features indicating the slab track conditions. Surprisingly, WPT, STFT, Mel spectrogram, SST_STFT, and VMD are not able to classify the non-degraded state correctly. This is because the wavelet coefficients or spectrograms of Class 1 (No degradation) and Class 2 (Intermediate degradation) are quite similar. The WPT, STFT, Mel spectrogram, SST_STFT, and VMD fail to produce a higher frequency resolution in the low frequency range (dominant frequency range of slab track vibration). Therefore, they lack the ability of effectively distinguishing Class 1 and Class 2 in a robust way.

AG-ASDF shows a very good performance on all classes and outperforms other feature learning / extraction methods with an accuracy improvement between 8.0% and 46.6% compared to other methods. This indicates that AG-ASDF has the ability to learn effective thresholds for acoustic signal denoising and limits the impact of the speed on the obtained features and, thus, achieves a robust performance in the classification task for all the classes.

*2) Other compositions of training and test datasets*

We evaluate three other compositions of the training and test datasets as listed in Table IV.

TABLE IV
TRAIN SPEED ASSIGNMENT IN TRAINING AND TEST DATASET

| Composition Number | Training dataset | Test dataset |
|---|---|---|
| C1 | 40, 60, 80 km/h | 20 km/h |
| C2 | 20, 60, 80 km/h | 40 km/h |
| C3 | 20, 40, 80 km/h | 60 km/h |

Table V shows the average classification accuracy of classifying the three types of slab track under each composition of training and test datasets. The proposed AG-ASDF performs consistently as the best in all three conditions with an accuracy improvement between 1.8% and 40.7% compared to other feature extraction and learning methods. In addition, the classification tasks related to speed interpolation regime (dataset compositions C2 and C3) achieve a globally better classification performance compared to the tasks related to speed extrapolation regime (dataset composition C1 and the case described in Section C (1)).

TABLE III
COMPARATIVE RESULTS BASED ON ACOUSTIC SIGNALS IN CLASSIFICATION TASK 2

| Slab track condition | Classification accuracy (%) | | | | | | | | | |
|---|---|---|---|---|---|---|---|---|---|---|
| | AG-ASDF | DeSpaWN | FDWT | WPT | STFT | Mel | SST_STFT | SST_CWT | VMD | ST |
| No degradation (Class 1) | **98.6**±0.4 | 79.6±2.1 | 94.4 | 2.8 | 5.6 | 0.0 | 2.8 | 52.8 | 0.0 | 38.9 |
| Intermediate degradation (Class 2) | **85.6**±1.8 | 80.5±1.9 | 66.7 | 41.7 | 66.7 | 52.8 | 50.0 | 77.8 | 30.6 | 52.8 |
| Severe degradation (Class 3) | 86.1±1.2 | 86.1±1.6 | 83.3 | 100.0 | 100.0 | 100.0 | 100.0 | 100.0 | 100.0 | 100.0 |
| Average | **90.1** | 82.1 | 81.5 | 48.2 | 57.4 | 50.9 | 50.9 | 76.9 | 43.5 | 63.9 |

TABLE V
AVERAGE CLASSIFICATION ACCURACY BASED ON ACOUSTIC SIGNALS

| Dataset composition | Classification accuracy (%) | | | | | | | | | |
|---|---|---|---|---|---|---|---|---|---|---|
| | AG-ASDF | DeSpaWN | FDWT | WPT | STFT | Mel | SST_STFT | SST_CWT | VMD | ST |
| C1 | **57.4**±6.7 | 49.6±6.2 | 36.1 | 40.7 | 46.7 | 39.8 | 55.6 | 38.0 | 38.0 | 33.3 |
| C2 | **94.4**±3.0 | 83.4±5.2 | 79.6 | 72.2 | 74.1 | 76.8 | 88.0 | 75.9 | 53.7 | 75.9 |
| C3 | **94.7**±4.2 | 85.2±4.8 | 84.3 | 67.6 | 79.6 | 79.6 | 82.4 | 63.0 | 59.3 | 76.8 |

TABLE VI
AVERAGE CLASSIFICATION ACCURACY BASED ON AG-ASDF WITH DIFFERENT LOSS WEIGHTS (%)

| Training dataset | Test dataset | Weight ratio of reconstruction loss to guidance loss | | | | | | |
|---|---|---|---|---|---|---|---|---|
| | | 1:0 | 1:0.1 | 1:0.5 | 1:1 | 1:2 | 1:10 | 0:1 |
| 20, 40, 60, 80 km/h | 20, 40, 60, 80 km/h | 87.9 | 94.4±2.2 | 95.8±1.7 | 95.4±1.1 | **96.3**±1.4 | 95.8±1.1 | **96.3**±1.1 |
| 20, 40, 60 km/h | 80 km/h | 81.5 | 88.9±1.4 | 89.6±2.2 | 90.1±1.1 | **92.6**±3.1 | 92.1±1.9 | 92.1±1.8 |
| 40, 60, 80 km/h | 20 km/h | 36.1 | 51.2±8.1 | 53.2±3.4 | **57.4**±6.7 | 57.0±5.1 | 57.3±7.6 | **57.4**±8.6 |
| 20, 60, 80 km/h | 40 km/h | 79.6 | 93.1±4.6 | 93.5±5.3 | 94.4±3.0 | 94.4±2.9 | **94.5**±3.3 | 94.0±4.3 |
| 20, 40, 80 km/h | 60 km/h | 84.3 | 91.7±3.7 | 92.3±4.4 | **94.7**±4.2 | 94.4±4.4 | 93.5±4.2 | 94.0±2.5 |
| Average | | 73.9 | 83.9 | 84.9 | 86.4 | **86.9** | 86.6 | 86.8 |

It is worth noting that the classification accuracy on the dataset composition C1 based on acoustic signals is the lowest and even hardly reaches 60% accuracy. The test dataset of C1 is formed by acoustic signals with the train speed of 20 km/h. The slab track vibration under this slow train speed is weak, while some noise such as the machine noise produced by the train and the background noise are basically constant regardless of the train speed. Fig. 8 compares the magnitude of background noise and acoustic signals during the train passes with low and high speeds, which indicates a low SNR of the acoustic signal with the train speed of 20 km/h. The target features covered by severe noise cannot be effectively extracted by the current feature learning / extraction methods, which leads to the low classification accuracy in the C1 setup. Although the acoustic signal denoising based on AG-ASDF is not powerful enough in this case, the classification performance is still significantly improved by learning meaningful features with AG-ASDF compared to other feature extracting / learning approaches. Moreover, it is worth noting that this on the one hand the most difficult task due to a low SNR but it is also an extrapolation setup, since the algorithms are only trained on data collected with 40, 60 and 80 km/h. This explains the comparatively bad classification performance of all of the applied approaches.

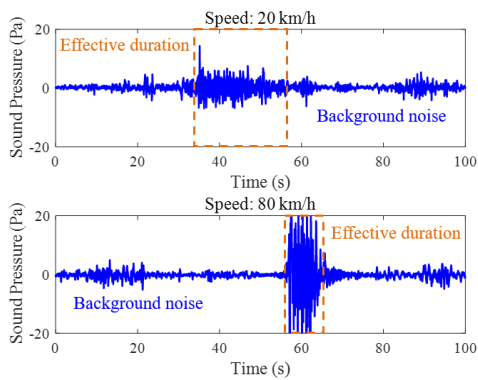

Fig. 8. Acoustic signals collected near Slab track 1 and under the train speeds of 20 and 80 km/h.

The comparison of classification results between Task 1 and Task 2 demonstrates the importance of composing a training dataset with data collected under sufficiently representative operating conditions for slab track state monitoring. Detecting the slab track condition under the train speed that exceeds the speed range in the training dataset (speed extrapolation regime) may result in a low detection accuracy.

### D. Loss weights of AG-ASDF

Reconstruction loss and guidance loss are two mutually exclusive loss terms in the process of acoustic signal denoising to enforce a moderate denoising result. The AG-ASDF with different loss weights is tested to analyze the influence of the two loss terms on the final classification performance. The average values of the classification accuracy in each condition monitoring task are listed in Table VI.

It appears from Table VI that the classification results vary with the weights of reconstruction and guidance loss terms in the AG-ASDF. The guidance loss with a too low weight reduces the performance of the AG-ASDF method to that of a FDWT which is not adapted for signal denoising (the case 1:0 corresponds to the FDWT). On average, increasing the weight of the guidance loss improves the classification performance of all condition monitoring tasks. However, the results are slightly impacted if the proportion of the guidance loss to the reconstruction loss reaches a certain threshold, which is judged as 1:1 as shown in Table VI. Considering that the classification results are roughly the same above this threshold of weight ratio, either 1:1 or 0:1 could be a good choice for the proportion of the reconstruction loss to the guidance loss.

## IV. CONCLUSION

In this study, we use three types of slab track with different supporting conditions (mortar, rubber and spring support) in a railway test line to imitate different healthy and unhealthy states of slab tracks. Track-side acoustic and acceleration signals are collected as a train passes by with different speeds. We propose a framework, called AG-ASDF, that can automatically denoise acoustic signals and learn meaningful representations under the guidance of the corresponding acceleration signals during the training step. We compared the performance of our framework to several other feature learning and extraction methods on two classification tasks (Task 1 comprises of the same train speeds in training and test dataset, Task 2 comprises different train speeds in training and test dataset), partly with extrapolation setups.

For both tasks, the proposed AG-ASDF framework reaches a superior performance compared to other feature extraction and learning methods. Besides, equipped with flexible time and frequency resolutions at different decomposition layers (which are the inherent characteristics of the DeSpaWN algorithm), AG-ASDF has the ability to embed the acoustic signal into a relevant representation which is robust to noise and different train speeds in the classification task.

Because the effect of acoustic signal denoising mainly depends on the guidance loss instead of the reconstruction loss in the AG-ASDF, the guidance loss contributes more to the performance improvement than the reconstruction loss for the classification tasks in this study. Increasing the weight of the guidance loss can improve the classification performance whereas the results are influenced slightly if the proportion of guidance loss to reconstruction loss rises continuously after reaching a certain threshold. It indicates that either 1:1 or 0:1 could be a good choice for the proportion of the reconstruction loss to the guidance loss in the case of using our experimental data. Further investigation needs to be made to assess the quality of the acoustic reconstruction with AG-ASDF under different acoustic monitoring scenarios.

Generally, from the performed experiments it appears that speed interpolation regimes perform better than speed extrapolation regimes in classification tasks. In particular, the classification performance of extrapolation to lower speed regime is considerably poor because of the low SNR of acoustic signals under the condition of 20 km/h. These results demonstrate the importance of collecting representative training datasets covering all relevant operating conditions.

In future work, additional evaluations are required to test the robustness and generalization ability of the proposed AG-ASDF with different train types, different positions of sensors, etc. Besides, the AG-ASDF will be evaluated on other acoustic monitoring tasks such as bearing condition monitoring, bolt looseness identification, and rail defect inspection to promote the extensive application of the acoustic denoising framework.